\documentclass[11pt,twoside]{article}
\usepackage{a4,supertabular,lscape,color,newpasp,subfigure}
\input{psfig.sty}

\begin{document}

\title{Deep ROSAT-HRI observation of the elliptical galaxy NGC 1399}

\author{M. Paolillo$^1$, G. Fabbiano$^2$, G. Peres$^1$, D.-W. Kim$^2$}
\affil{$^1$ Universit\`a di Palermo, D.S.F.A. - Sez.di Astronomia, Piazza del Parlamento 1,90134 Palermo, Italy\\
$^2$ Harvard-Smithsonian Center for Astrophysics, High Energy Astrophysical Division, 60 Garden St., Cambridge, Massachusetts 02138\\}

\begin{abstract}
We present the preliminary results of a deep (167 ks) ROSAT HRI
observation of the cD galaxy NGC1399 in the Fornax cluster. We find, in agreement with previous observations, an extended (42-46 Kpc adopting a distance of 19 Mpc) gaseous halo with a luminosity of $L_X=(4.41\pm 0.04)\times10^{41}$ erg s$^{-1}$. The 5 arcsec resolution of the data allows us to detect a very complex and asymmetric structure of the halo with
respect to the optical galaxy. Moreover the analysis of the radial structure reveals the presence of a multi-component profile not consistent with a simple King model over the whole 40 Kpc. We do not detect the presence of a central source and  pose an upper limit to the luminosity of a possible active nucleus.

Due to the length of the observation, comparable to that of a deep survey,
we detect a large number of sources within the HRI FOV, in slight excess with respect to the estimates based on previous surveys. We study the flux distribution of the sources, their temporal behaviour and their spatial distribution with respect to the central galaxy.
\end{abstract}

\section{Observations}
\label{Composite observation}
The NGC 1399 field, including NGC 1404, was observed at three separate times
with the ROSAT High Resolution Camera (HRI): in February 1993, 
between January and February 1996 and between July and August of the same year
(Table \ref{obs_tab}).
The total exposure time is 167567 s.

\begin{figure}[!t]
\centerline{\psfig{figure=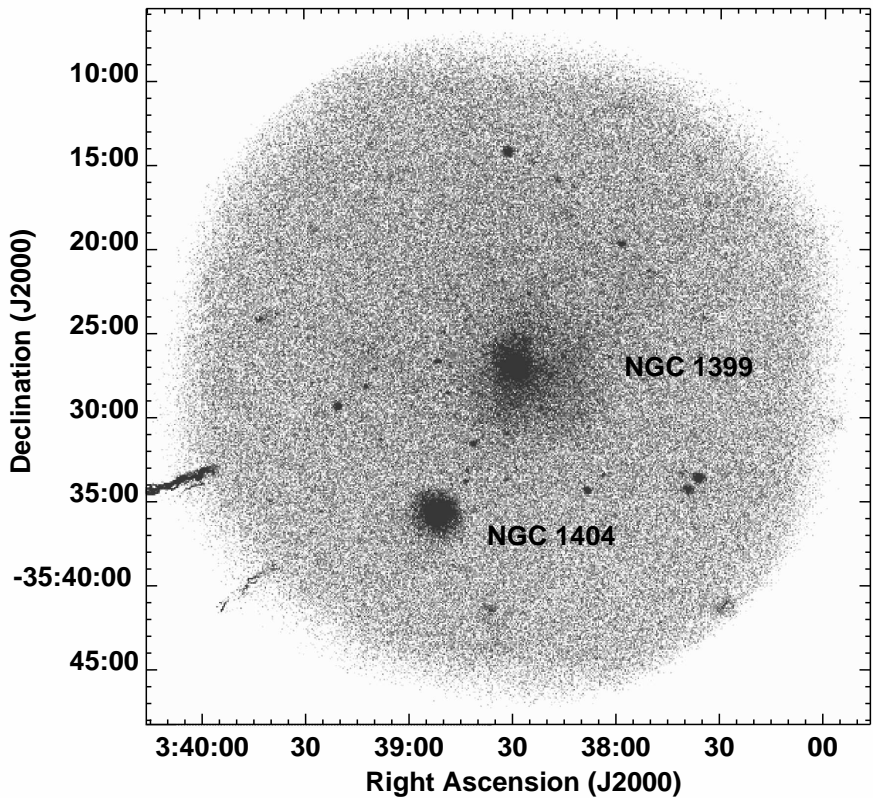,angle=0,width=1.1\textwidth}}
\caption{The raw NGC 1399/NGC 1404 field. The composite image, obtained adding the three observations of Table \ref{obs_tab}, was rebinned to 5 arcsec/pixel. The two brightest sources represent NGC 1399 (center of FOV) and NGC 1404 (South-East prominent source); even without exposure correction the extended emission sourrounding NGC 1399 is clearly visible. Many additional sources (see Fig.\ref{det}) are present in the field. Their properties are listed in Table \ref{sources_tab}
The elongated features in the lower left
corner are due to the presence of ``hot spots'' on the detector.}
\label{NGC 1399comp}
\end{figure}

The observations were corrected for aspect problems (Harris et al. \cite{Harris98}) and added together to obtain a ``{\bf composite
observation}''. To correct pointing uncertainties, that may affect the following analysis, we used the centroids of three bright pointlike sources in the field
(N.8, 19 and 24 in Table \ref{sources_tab}) to align the observations.  
The final image is shown in Figure \ref{NGC 1399comp}.

\begin{table}[h]
\caption{Rosat HRI observations of NGC 1399}
\label{obs_tab}	
~\\
\centerline{
\begin{tabular}{ccccrcc}
\hline									
Obj.name & \multicolumn{2}{c}{Field center} & sequence id. &
 Exp.time & obs. date & P.I.\\
& ra & dec & & (sec) & &\\
\hline
NGC 1399  & 03:38:31 & --35:27:00 & RH600256n00 & 7265 & 17/02/93 & D.-W. Kim\\
''  & '' & '' & RH600831n00 & 72720 & 04/01/96-23/02/96 & G. Fabbiano\\
''  & '' & '' & RH600831a01 & 87582 & 07/07/96-26/08/96 & G. Fabbiano\\
\hline
\end{tabular}}
\end{table}

\section{Brightness distribution and X-optical comparison}
\label{brightness}
To study the brightness distribution of the NGC 1399 field we rebinned the data in 5''$\times$5'' pixels and adaptively smoothed the image with the CSMOOTH algorithm contained in the CIAO package, developed at the Smithsonian Astrophysical Observatory. The algorithm convolves the data with a gaussian of variable width so as to enhance, at the same time, both small and large scale structures.

\begin{figure}[p]
\centerline{\subfigure{\psfig{figure=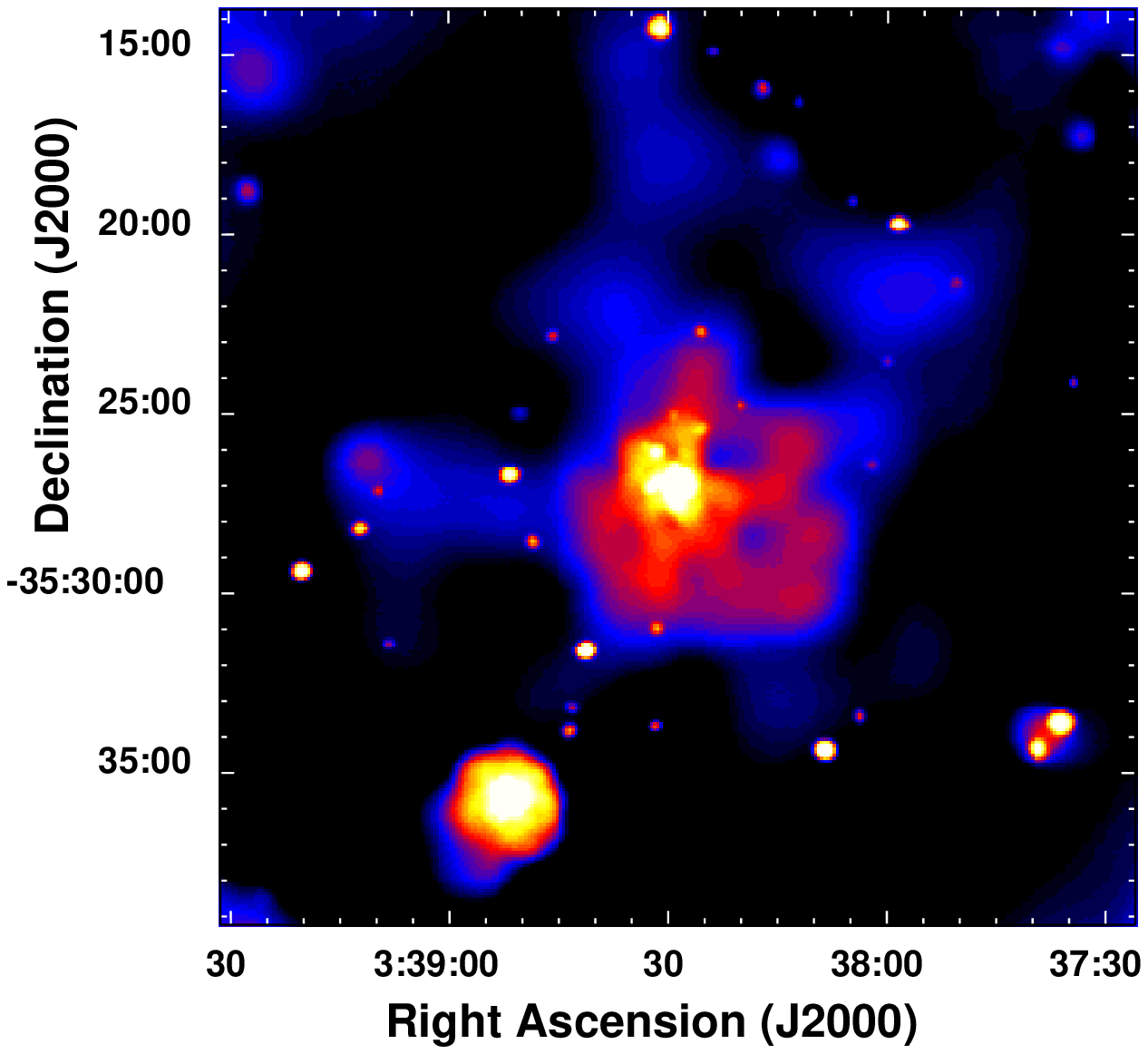,angle=0,height=0.4\textheight}}}
\centerline{\subfigure{\psfig{figure=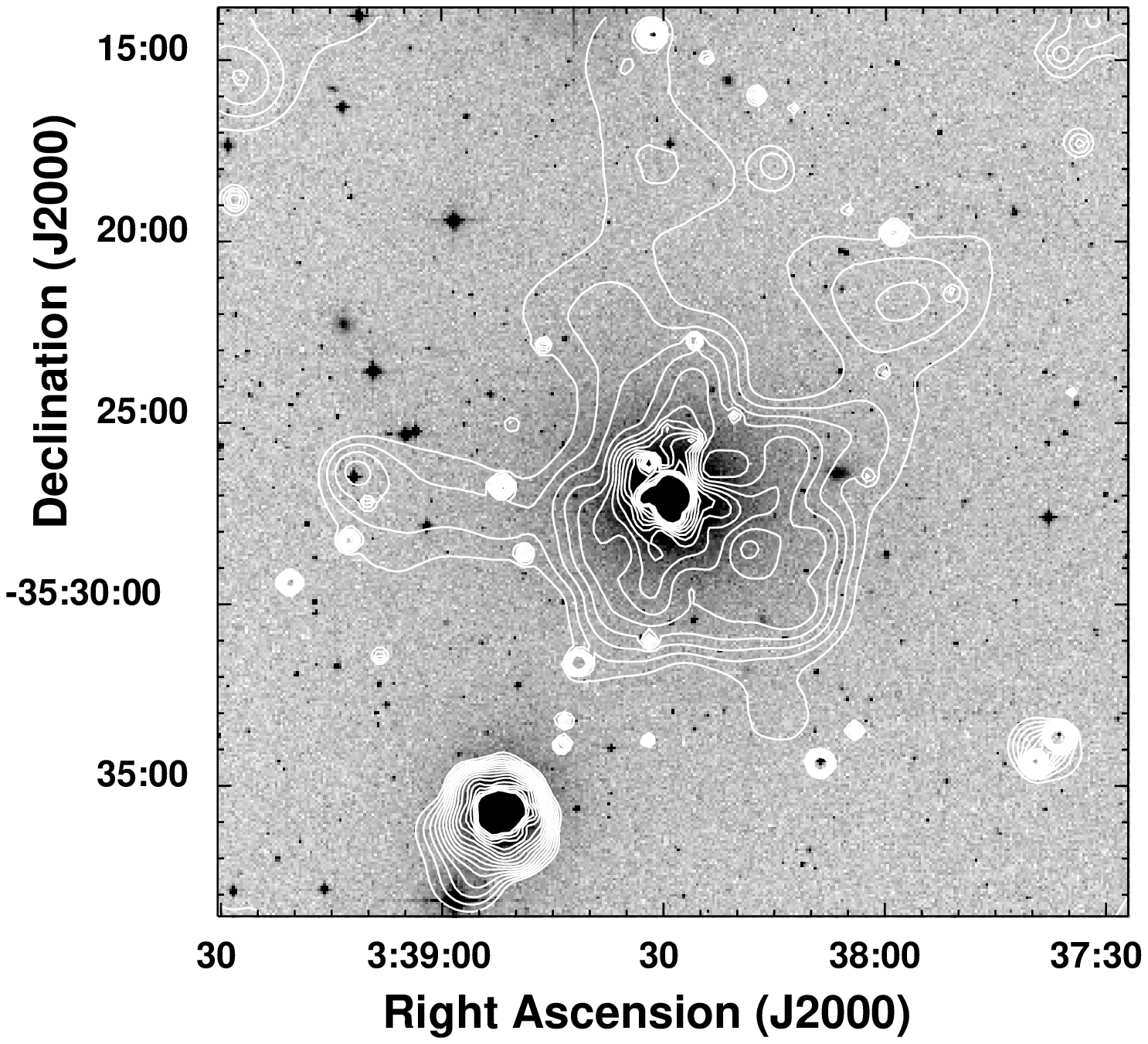,angle=0,height=0.4\textheight}}}
\caption{{\bf Upper panel:} Adaptively smoothed image of the NGC 1399/1404 field. Colors from black to yellow represent X-ray intensities from $5.6\times 10^{-3}$ to $3.58\times 10^{-1}$ erg arcmin$^{-1}$ s$^{-1}$. The image is binned
at $5\times 5$ arcsec/pixel resolution. The region corresponds to the yellow box in Figure \ref{NGC 1399comp}. {\bf Lower panel:} The X-ray brightness contours overlaid on the 1 arcsec/pixel DSS image (logarithmic grayscale). Contours are spaced by a factor 1.1 with the lowest one at $6.1\times 10^{-3}$ erg arcmin$^{-1}$ s$^{-1}$.} 
\label{opt_contours}
\end{figure}

The "csmoothed" (and exposure corrected, Snowden et al. \cite{Snow94}) 5"/pixel image shown in Figure \ref{opt_contours}, reveals a very complex X-ray morphology.
The center of the image is occupied by the extended halo of NGC 1399. 
The galaxy has a central compact component within $\sim 1$ arcmin from the X-ray emission peak and an external extended and asimmetric halo
The halo is decentered with respect to the X-ray peak, extending more on the SW side. Inside the halo we see many elongated structures protruding from the nuclear region separated by ``voids''. Over 5 arcmin from the center the halo seems to replicate on larger scale the features seen in the inner 5 arcmin, extending further with larger elongated structures and  voids.
As it can be seen from a comparison with the optical DSS image (Fig.\ref{opt_contours} - lower panel),
while the X-ray emission peak is centered on the optical galaxy, the halo extends much further than the optical distribution. Moreover most of the features seen in the X-ray image have no optical counterpart and must then be part of NGC 1399 halo.

A comparison with NGC 1404, whose X-ray emission is almost symmetric and consistent with the optical distribution, confirms the peculiar nature of dominant cluster members such as NGC 1399. The "csmoothed" image reveals also the presence of the SE tidal tail already seen by Jones et al.(\cite{jones97}, hereafter JSF) in PSPC data, probably a signature of ram pressure stripping of the NGC 1404 corona due to the infall toward the NGC 1399 halo.

The great number of pointlike sources present in the field is also evident in Figure \ref{opt_contours}. These objects will be studied in greater detail in Sec.\ref{sources}

\section{Radial brightness profiles}
\label{profiles}

\subsection{NGC 1399}
\label{NGC 1399}
\begin{figure}[!t]
\centerline{\psfig{figure=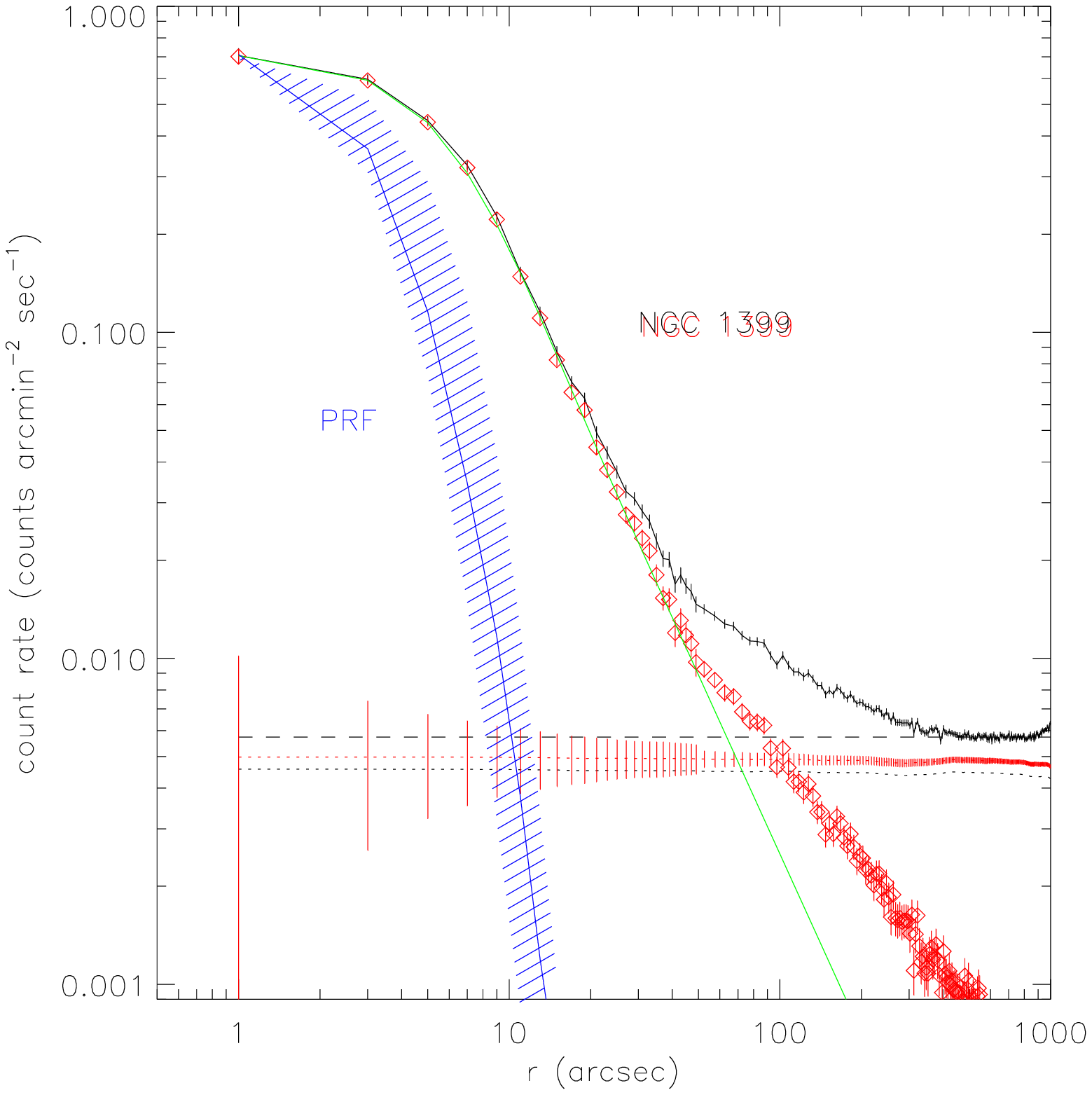,angle=0,width=0.4\textheight}
}
\caption{Radial profile of NGC 1399 X-ray surface brightness (black continuous line) after excluding all point sources reported in Table \ref{sources_tab} and NGC 1404. The long-dashed line represents the flattening level of the HRI profile measured in the 500''-850'' annulus.
Black and red dotted lines represent respectively the particle background level before and after rescaling to match PSPC counts (see text).
The rescaled background has been used to derive the background-subtracted counts (red diamonds) and the best-fit Beta model within 50"(green line). 
The HRI PRF range is represented by the blue shaded region. Bins are 2" wide up to r=50" and 5" wide at larger radii.}
\label{profile}
\end{figure}

To study the brightness distribution of NGC 1399 we created the radial brightness profile shown in Figure \ref{profile}. The emission extends out to at least 450''--500'', equivalent to 42--46 Kpc (1 arcmin = 5.53 Kpc for a distance of 19 Mpc, H$_0$=75 Km s$^{-1}$/Mpc). We know from previous investigations of the Fornax cluster (JSF, Ikebe et al. \cite{ikebe96}) that extended X-ray emission is present over 500''. Unfortunately, over this radius, the HRI data are affected from exposure correction uncertainties so that we can not clearly detect the outer X-ray emission.
   
After comparing our data with the Rosat-PSPC profile we adopted as background
level (red dotted line in Figure \ref{profile}) the rescaled HRI particle map produced with the Snowden software (Snowden et al. \cite{Snow94}; for a more detailed discussion of the background determination see Paolillo et al., \cite{Paolillo}). This background level was used to derive the background subtracted profile shown as red diamonds in Figure \ref{profile}. We can discriminate an {\bf inner component}, decreasing with a power-law profile, dominating up to 50'' and already evident in the brightness map of  Figure \ref{opt_contours}, plus an {\bf external component}, dominating over 50'', with a more extended profile. A comparison with the HRI point response function (blue shaded region) shows that both components are extended.

\begin{figure}[t]
\centerline{\psfig{figure=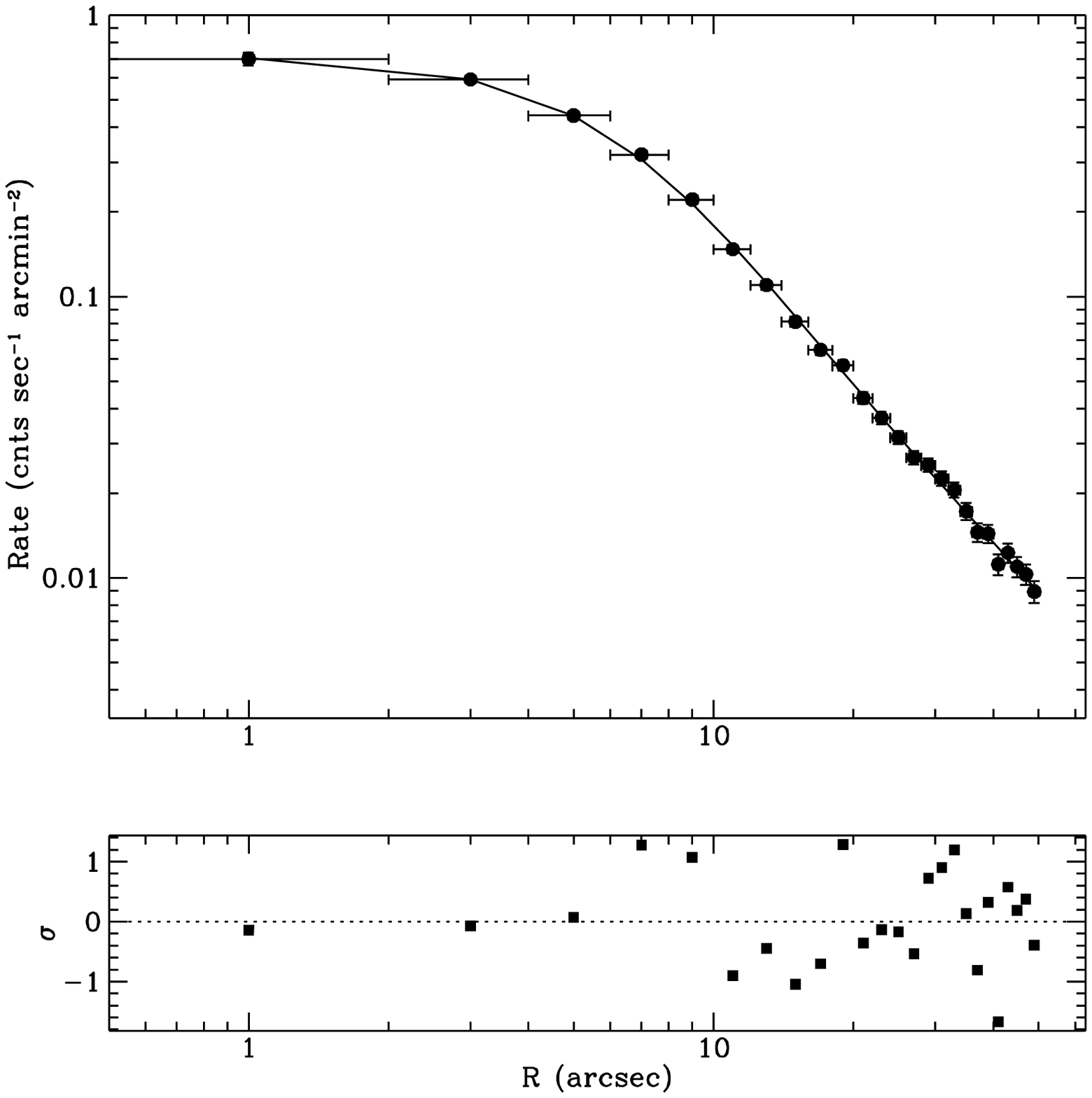,angle=0,width=0.6\textwidth}
\psfig{figure=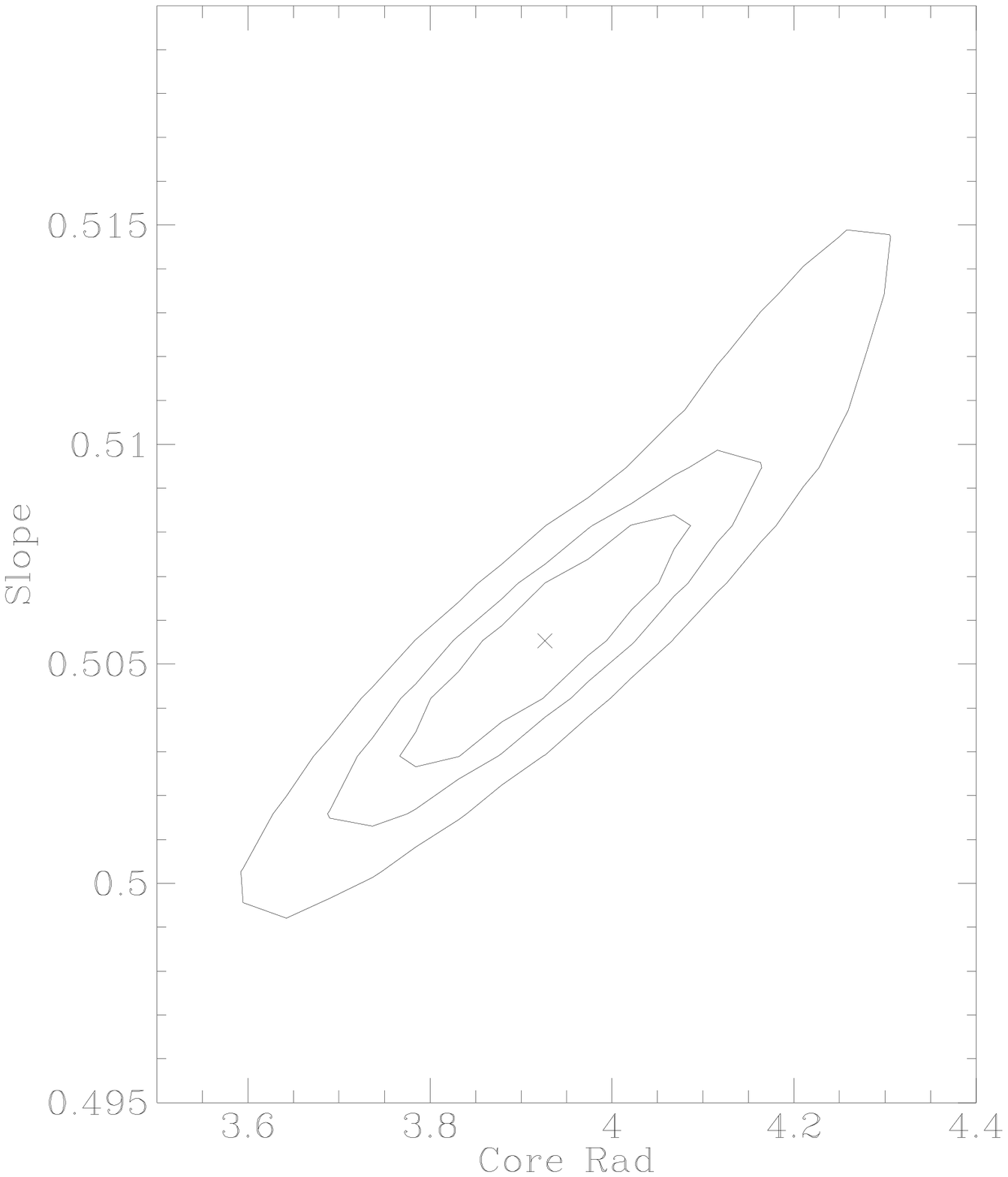,angle=0,width=0.55\textwidth}}
\caption{{\bf Left panel:} best-fit Beta model and residuals within 50'' of NGC 1399 brightness profile, obtained binning the data in 2'' annuli. {\bf Right panel:} 66\%, 90\% and 99\% contour levels relative to the the model parameters.}
\label{fit}
\end{figure}
\begin{figure}[t]
\centerline{\psfig{figure=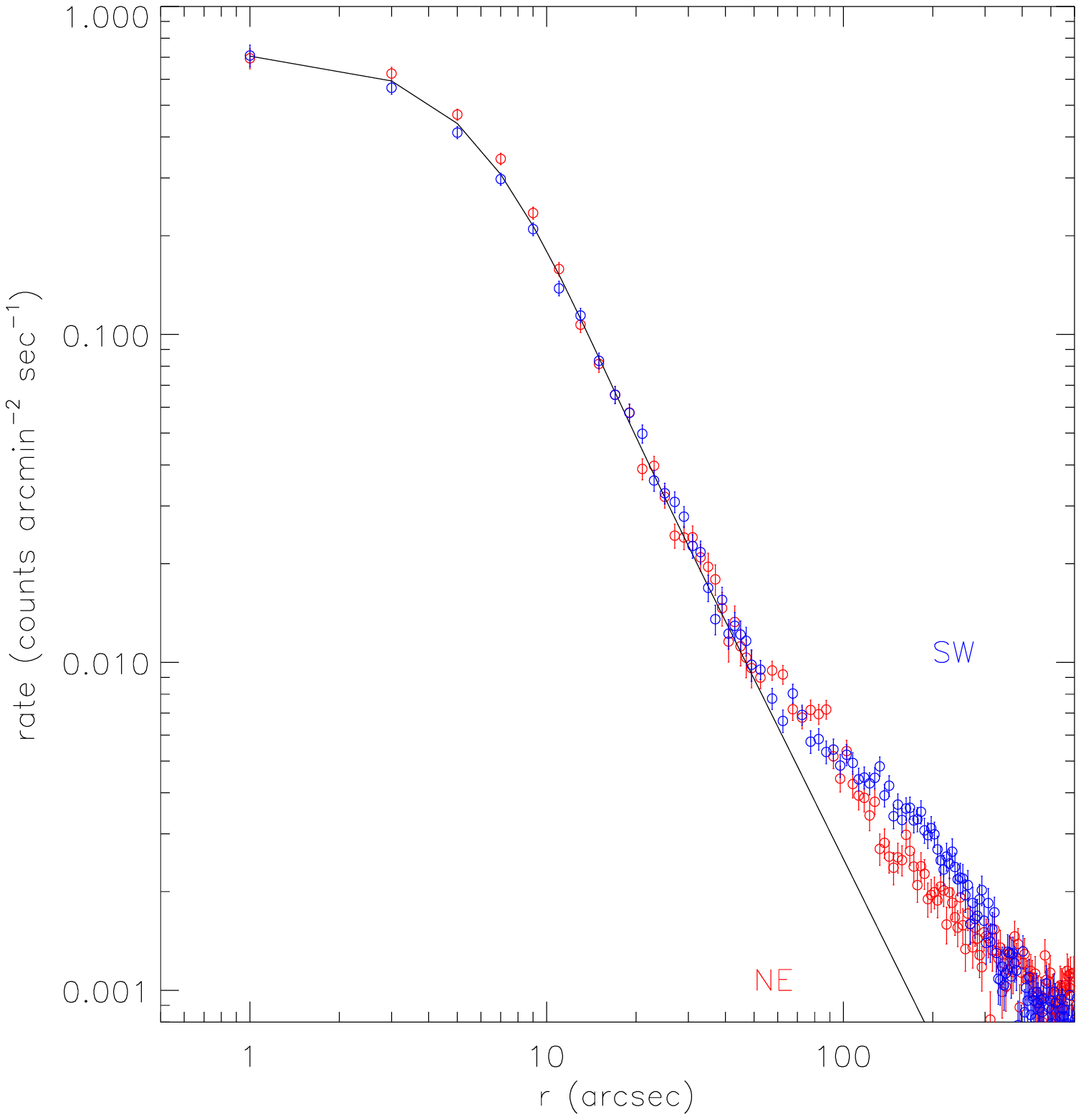,angle=0,width=0.65\textwidth}}
\caption{Background subtracted radial profile of NGC 1399 X-ray surface brightness, binned in 5'' annuli, in the NE and the SW sectors. The best-fit model within the inner 50'' is shown as a continuous line.}
\label{pies}
\end{figure}

Due to the complex halo structure it is not possible to model the whole brightness distribution with a single model. We thus decided to model the inner component, up to 50'', with the Beta profile:
$$\Sigma\propto [1+(r/a)^2]^{-3\beta+0.5}$$
binning count rates in 2'' annuli and properly convolving for the HRI PRF.
The best-fit model is compared the observed profile in Figure \ref{fit}. We obtain the best-fit values 
$$a=3.93\pm 0.16~~~~~~~\beta=0.506\pm 0.003$$
yielding $\chi^2=14.5$ for 22 degrees of freedom. Thus, for $r >> a$ the
X-ray emission of the inner component falls as $r^{-2.04\pm 0.02}$. 

To study the outer halo structure we derived the X-ray profiles in two different sectors. We divided our F.O.V. along the line connecting the NGC 1399 and NGC 1404 X-ray centroids: a NE sector (P.A. 331.5$^{\circ}$--151.5$^{\circ}$) and a SW sector (P.A. 151.5$^{\circ}$--331.5$^{\circ}$).
The radial profiles are plotted in Figure \ref{pies}. While the compact component shows minor asymmetries, the outer halo is more extended in the SW direction. Moreover in both sectors the X-ray emission is well in excess of the Beta profile derived from the best-fit of the inner 50'' (black continuous line).

The asymmetries of the NGC 1399 X-ray emission were already noticed by JSF. They showed that the outer halo extends at least out to 150 Kpc but, even though they find a central excess with respect to their best-fit model, their data do not have enough resolution to clearly separate the compact component from the outer one, thus obtaining smaller values for the `mean' power law slope ($\sim$0.35).  

We are currently exploring the possibility of modeling the whole halo with a double Beta profile to determine the relative contribution of the two components.

\subsection{NGC 1404}
\label{NGC 1404}
\begin{figure}[!h]
\centerline{\psfig{figure=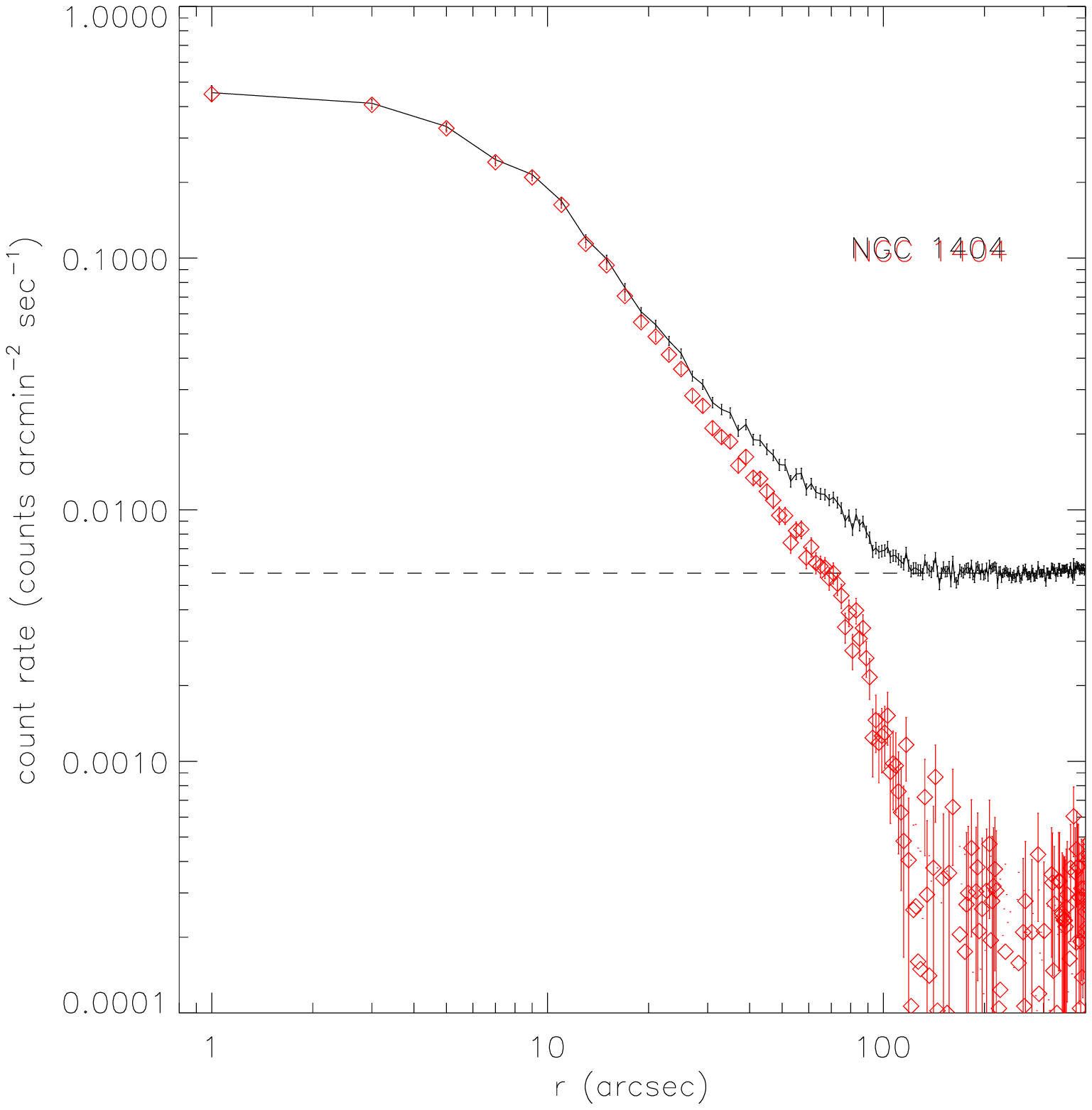,angle=0,width=0.65\textwidth}}
\caption{Radial profile of NGC 1404 extracted in 2'' annuli (black continuous line). The red data points represents the background subtracted profile using the count rate measured in the 200''-300'' interval (dashed line). The brightness distribution follows quite well a Beta profile up to $\sim 100''$, showing no sign of a multi-component structure as the one seen in NGC 1399.}
\label{1404profile}
\centerline{\psfig{figure=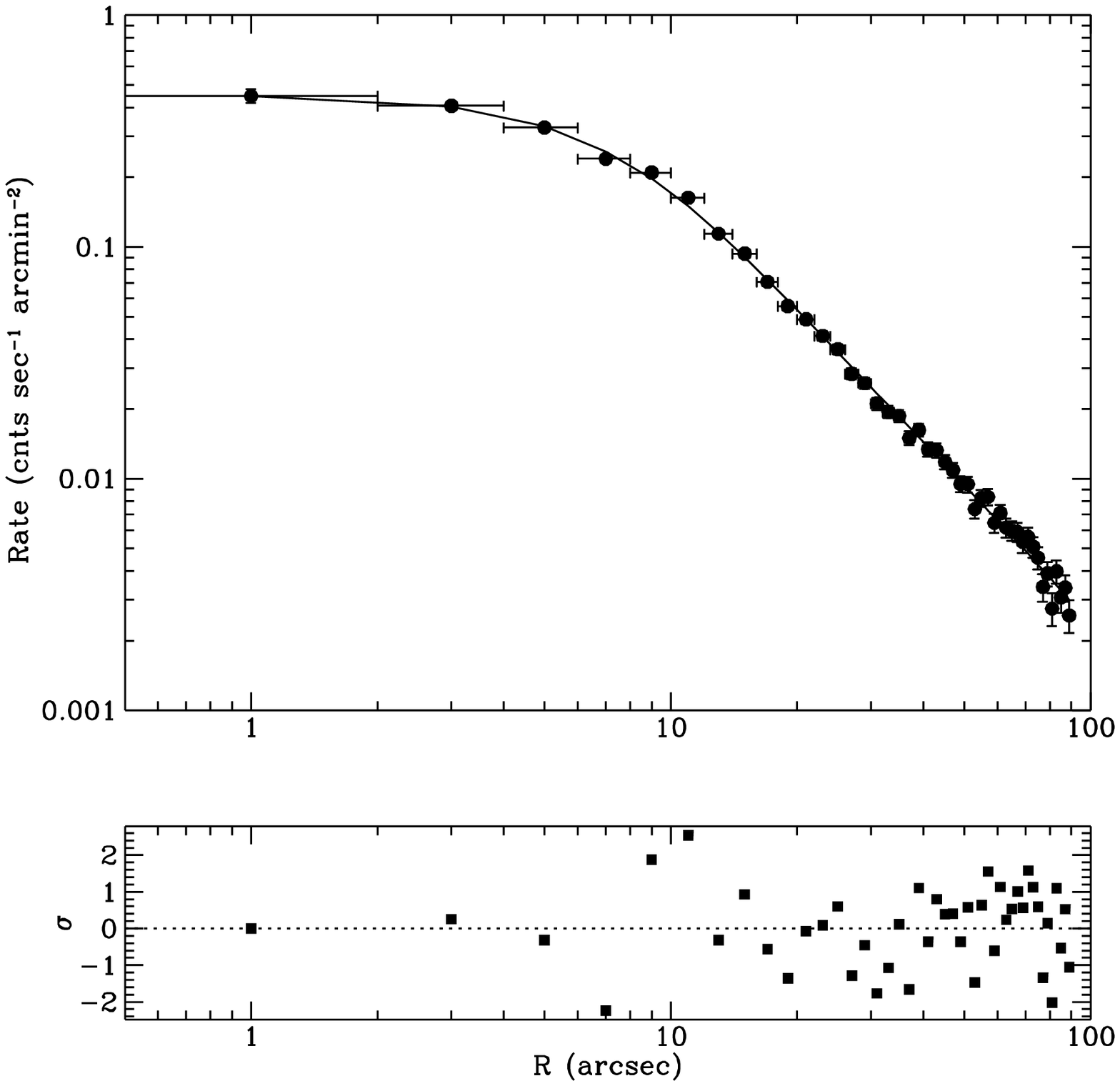,angle=0,width=0.55\textwidth}
\psfig{figure=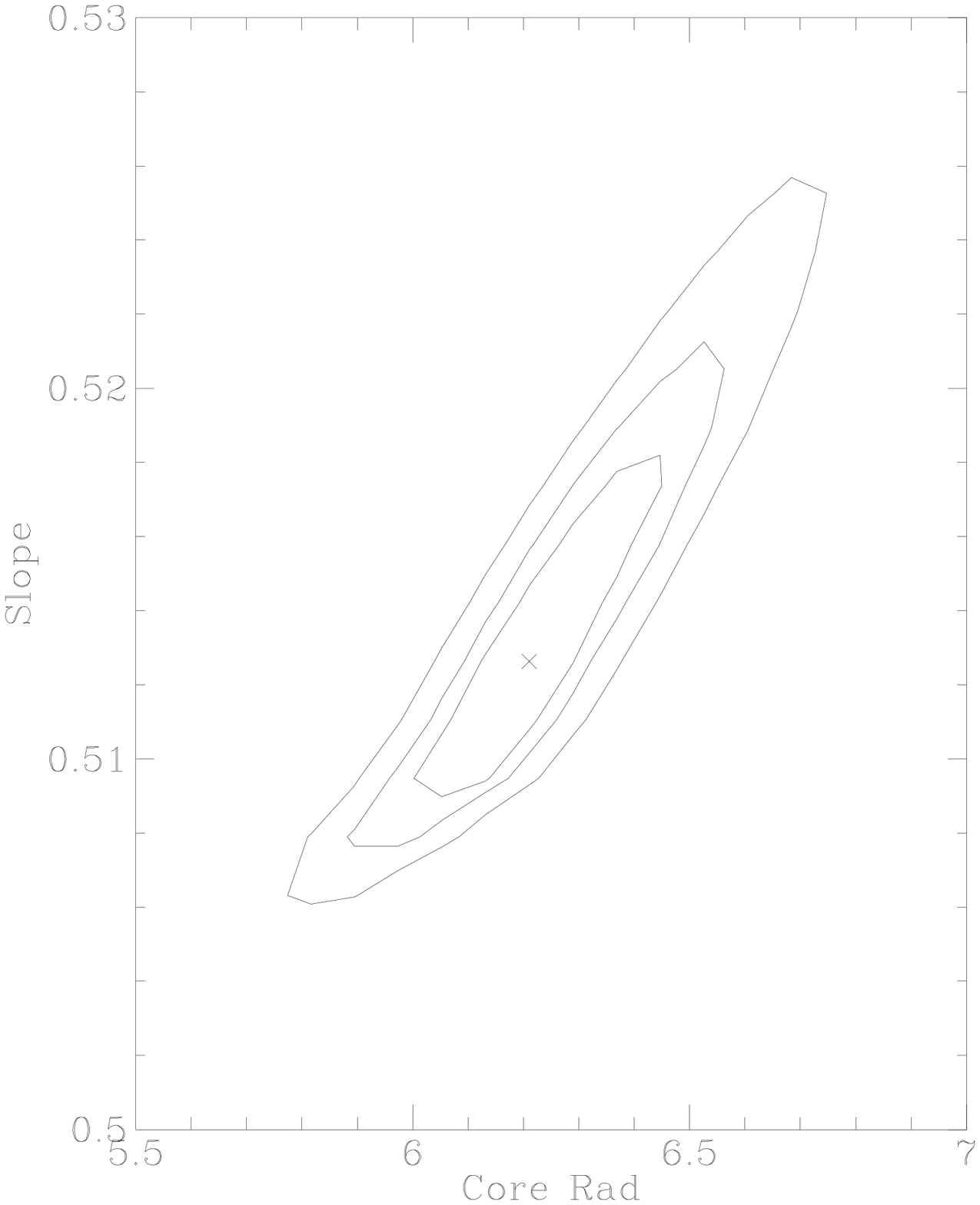,angle=0,width=0.52\textwidth}}
\caption{{\bf Left panel:} best-fit Beta model and residuals within 90'' of NGC 1404 brightness profile, obtained binning the data in 2'' annuli. {\bf Right panel:} 66\%, 90\% and 99\% contour levels relative to the model parameters.}
\label{1404fit}
\end{figure}

We derived the brightness profile of NGC 1404 extracting count rates in 2'' annuli centered on the galaxy centroid. A background level of $5.60\times 10^{-3}$ counts arcmin$^2$ sec$^{-1}$ counts was determined in a 200''-- 300'' annulus centered on the galaxy. The measured count rates are shown in  Figure \ref{1404profile}.
NGC 1404 emission extends up to 110'' ($\sim$10 Kpc) from the galaxy center and shows no sign of a multi component structure as the one detected in NGC 1399. The emission exibits a circular symmetry and declines with a Beta profile up to 100''.

We fitted the observed brightness distribution, within 90'', with a Beta model. 
obtaining:
$$a=6.21^{+0.25}_{-0.20}~~~~~~~~~\beta=0.513^{+0.005}_{-0.004}$$
with $\chi^2=51.5$ for 42 d.o.f.
For $r >> a$ the X-ray emission falls as $r^{-2.08\pm 0.03}$. 
The best fit model is shown in Figure \ref{1404fit} together with the  66\%, 90\% and 99\% contour levels relative to the fit parameters.

\section{Fluxes and Luminosities}
\label{fluxes}

NGC 1399 X-ray counts were extracted within 500'' and corrected for the contamination of sources listed in Table \ref{sources_tab}. Background counts were measured from the rescaled particle map (see Sec.\ref{NGC 1399}).
The total count rate is 0.451$\pm$0.004 counts sec$^{-1}$.

To obtain a flux estimate we used the two temperature Raymond-Smith model found by Rangarajan et al.(\cite{rang95}, hereafter RFFJ) using ROSAT PSPC data: a main thermal component with  $KT\simeq$1 keV plus a softer emission with $KT=80.8$ eV contributing for 16\% of the emission in the 0.2--0.3 keV range. The metal abundace was fixed to the solar value and the absorbing column to the galactic value ($1.3 \times 10^{20}$ cm$^{-2}$). We obtain $f_X=(1.14\pm 0.01)\times 10^{-11}$ erg s$^{-1}$ cm$^{-2}$ in the 0.1-2.4 keV energy range. Assuming a distance of 19 Mpc $L_X=(4.41\pm 0.04)\times10^{41}$ erg s$^{-1}$. Taking into account differences in the background subtraction and exposure correction uncertainties this result agrees within 10\% with the estimate of RFFJ.

\clearpage
In the case of NGC 1404 we used a single temperature Raymond-Smith spectrum with $KT=0.6$ keV and metallicity 0.2 of the solar abundance, following JSF. We measured a count rate of $0.277\pm0.003$ counts sec$^{-1}$ within 150'' extracting background from the 200''-300'' annulus, consistent within 7\% with the one derived by JSF. This gives a total flux $f_X=(3.64\pm 0.03)\times 10^{-12}$ erg s$^{-1}$ cm$^{-2}$ and a total luminosity of $L_X=(1.41\pm 0.01)\times10^{41}$ erg s$^{-1}$.

NGC 1399 is known to be a radio galaxy with a faint radio core ($S_{core}\sim 10$ mJy at 5 GHz, Ekers et al. \cite{ekers89}). Searching the X-ray counterpart of the radio core, we estimated an upper limit for a central point source: we added a central point source to our Beta model repeating the fit at different brightness values until we reached a $3\sigma$ confidence level. We obtain an upper count rate limit of $2.25\times 10^{-3}$ counts s$^{-1}$. Assuming a power law spectrum with $\alpha_{ph}=1.7$ we get $f_X^{3\sigma}=1.00\times 10^{-13}$ erg s$^{-1}$ cm$^{-2}$ in the 0.1-2.4 keV range, corresponding to $L_X^{3\sigma}=3.87\times 10^{39}$ erg s$^{-1}$. This value is approximately 50\% of what found in the spectral analysis of RFFJ.

\section{Individual sources}
\label{sources}
Our field contains many emission peaks in addition to the two main galaxies. We made use of the algorithm developed at the Palermo Observatory by F.Damiani and collaborators (Damiani et al. \cite{Dam97a}) to detect these sources.
The wavelet algorithm found 43 sources in our field, 
using a signal to noise threshold of 4.65, so to assure a contamination of one/two spurious detection (F.Damiani: private communication, see also Damiani et al. \cite{Dam97b}).

\begin{figure}[p]
\centerline{\psfig{figure=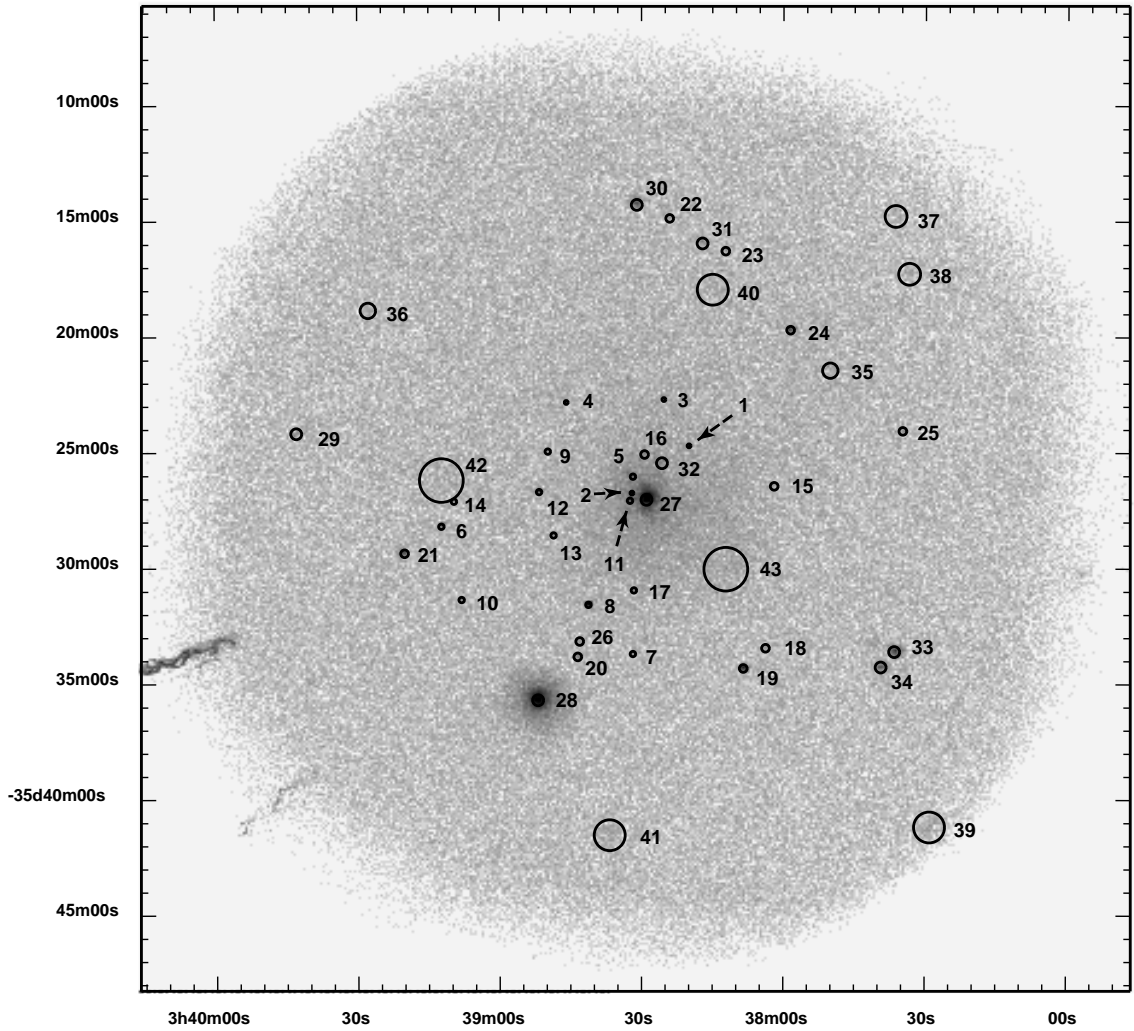,angle=0,width=1.3\textwidth}}
\caption{The sources detected by the Wavelets algorithm  of Damiani et
al.(\cite{Dam97a}, \cite{Dam97b}) superimposed on the 5''/pixel HRI FOV. Circles represent the scale of the maximum S/N ratio for each source.}
\label{det}
\end{figure}

Table \ref{sources_tab} lists the sources position, 3$\sigma$ radius, number of counts, the maximum signal to noise ratio within those measured by the algorithm at different scales, the count rate, the temporal variability based on a Kolmogorov-Smirnov test, the presence of a possible optical counterpart (identified by visual inspection of DSS photographic plates) and the name of the possible NED (Nasa Extragalactic Database) counterpart.  Sources N.27 and N.28, representing respectively NGC 1399 and NGC 1404, are not included in the table.

We compared our source distribution to the one obtained by Hasinger et al.(\cite{Has93}) from a deep survey of the Lockmann hole, after correcting for variable sensitivity across the detector.   
We assumed for our sources a power law spectrum with photon index $1.96\pm0.11$, as found by Hasinger for faint sources. Following what found by Kim \& Fabbiano (\cite{kim95}) in their study of pointlike sources near NGC 507, we also tried to use a Raymond-Smith spectrum with $KT=0.52$ and abundance $\sim 0.2$, obtaining an almost identical result.

\begin{figure}[!h]
\centerline{\psfig{figure=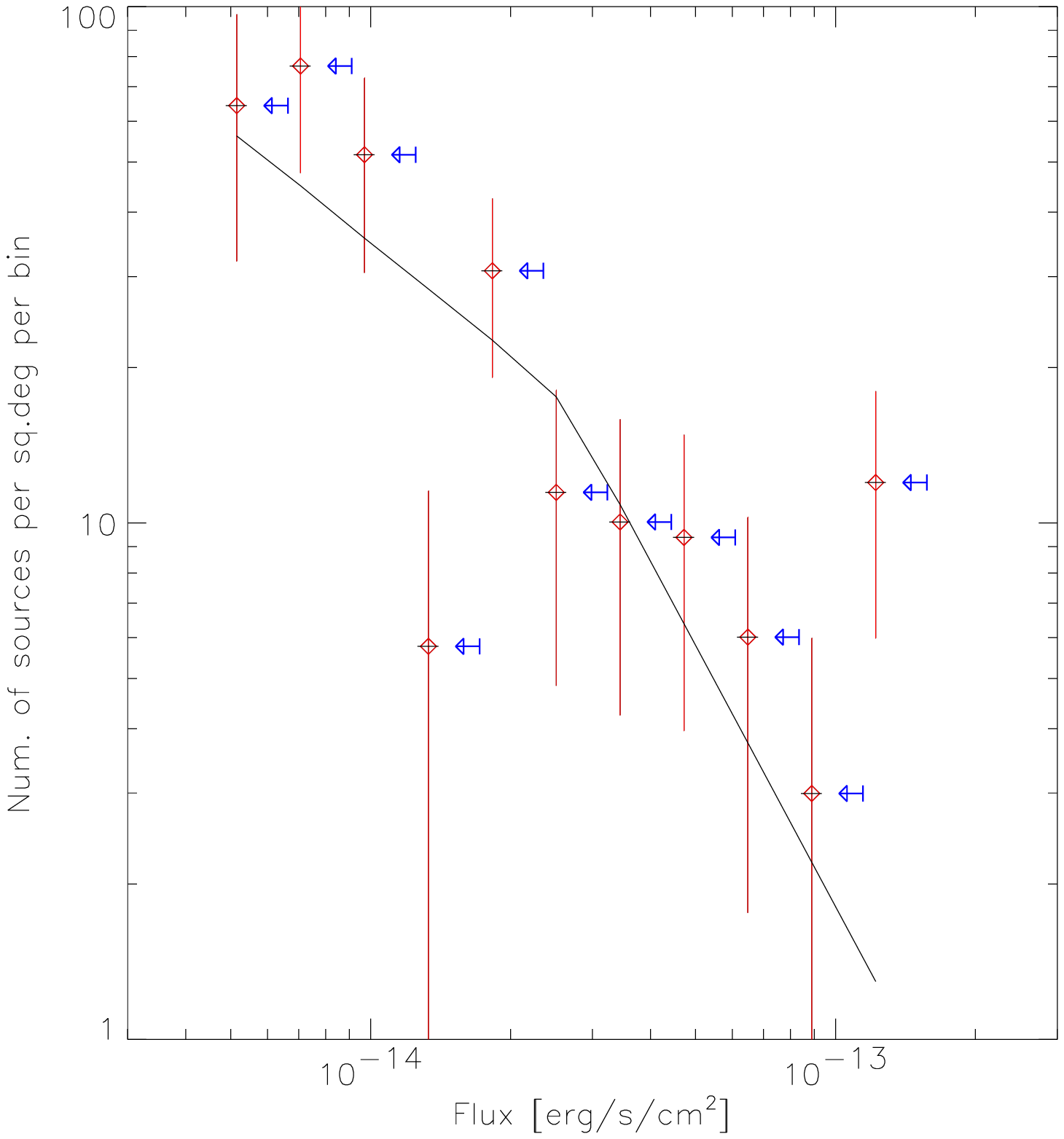,angle=0,width=0.38\textheight}}
\caption{Measured vs expected source counts. Measured counts, corrected for sensitivity variations across the detector, are represented by red points, while the black continuous line is the expected distribution from Hasinger et al.(\cite{Has93}). Horizontal error bars take into account uncertainties in the power law slope. Blue arrows mark the upper limits obtained fixing the absorption to the maximum value of $3.2\times 10^{20}$ allowed from the spectral analysis performed by Rangarajan et al.(\cite{rang95}).}
\label{src_flux}
\centerline{\psfig{figure=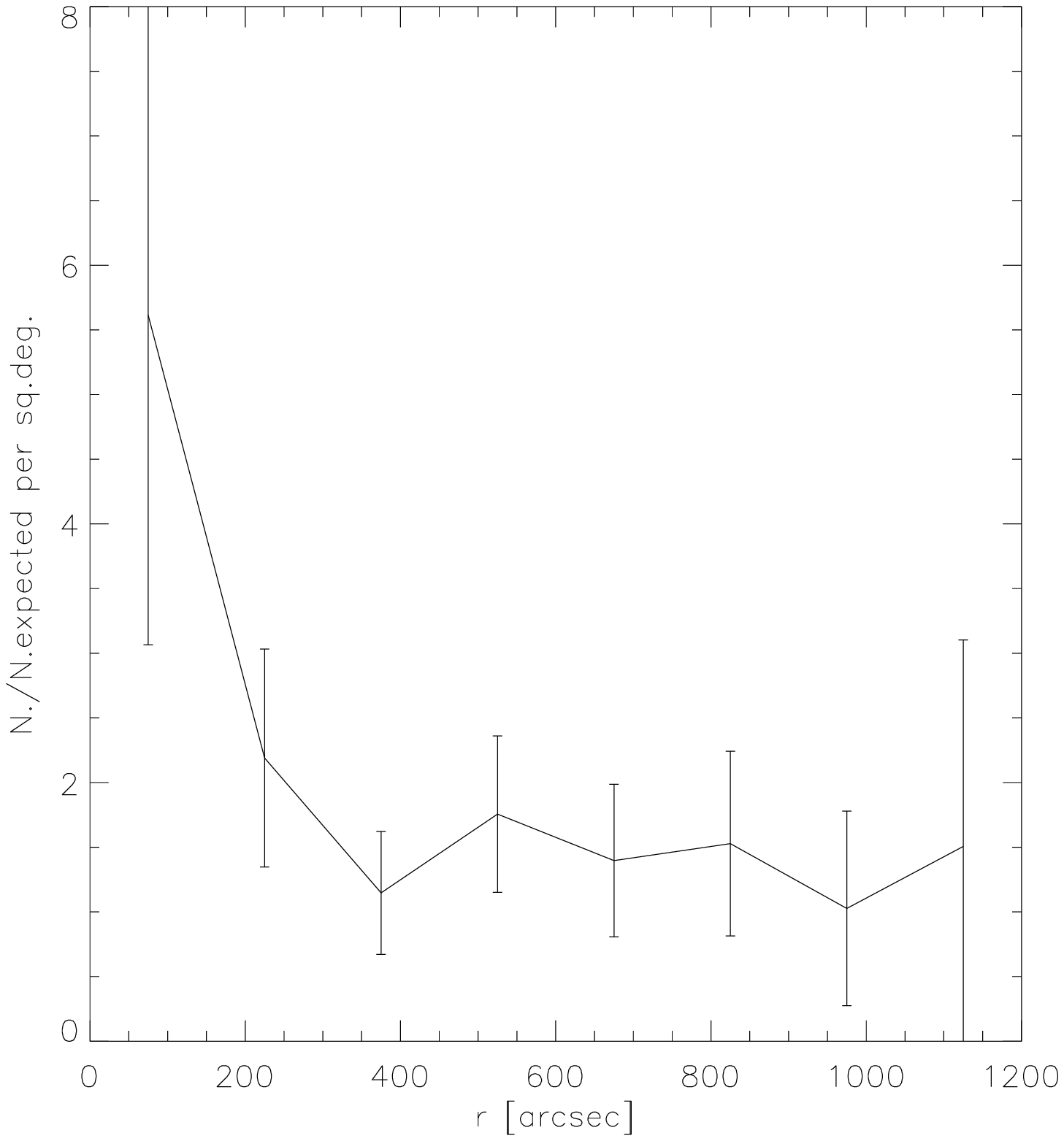,angle=0,width=0.38\textheight}}
\caption{Radial excess of sources over the expected number from Hasinger et al.(\cite{Has93}). The excess is concentrated around NGC 1399, thus suggesting to be due to sources associated with the galaxy.}
\label{src_excess}
\end{figure}

The resulting distribution is shown in Figure \ref{src_flux}. Our counts are higher than the Hasinger estimate but consistent at the $2\sigma$ level, giving a corrected total number of sources $281\pm 44$ over our field vs $230\pm 15$ expected counts. To take into account absorption uncertainties we estimated a lower limit on the number of expected sources adopting the higher absorption value found by RFFJ in the NGC 1399 field. In this case the number of expected counts decreases to $184\pm 14$ leading to an upper limit of the observed excess of $\sim 2.1\sigma$.

\begin{landscape}

\begin{table}[p]
\begin{supertabular}{cccrrrcccc}
\hline
N. & R.A.         & Dec.         & 3$\sigma$ Radius & Counts & Max S/N & Count Rate & Time Var.& Opt.id. & NED obj.\\
& & & (arcsec) & & & (Counts/sec) & (KS test) & & \\
\hline										
\\
1 & 03:38:20  & -35:24:42 & 8.66    & 35$\pm$12   & 4.65  & 2.1$\pm$0.7x10$^{-04}$ & - & - & - \\
	
2 & 03:38:32  & -35:26:45 & 8.66    & 54$\pm$17   & 6.10  & 3.2$\pm$1.0x10$^{-04}$ & - & - & - \\
	
3 & 03:38:25  & -35:22:42 & 8.66    & 49$\pm$14   & 6.67  & 2.9$\pm$0.8x10$^{-04}$ & - & - & - \\
	
4 & 03:38:46  & -35:22:50 & 8.66    & 42$\pm$12   &  6.09 & 2.6$\pm$0.7x10$^{-04}$ & - & y & CGF 0202\\
	
5 & 03:38:32  & -35:26:02 & 11.88   & 165$\pm$23  & 13.04 & 9.7$\pm$1.4x10$^{-04}$ & - & - & CGF 0233\\
	
6 & 03:39:12  & -35:28:12 & 13.23   & 172$\pm$21  & 14.91 & 1.0$\pm$0.1x10$^{-03}$ & - & y & - \\
	
7 & 03:38:32  & -35:33:42 & 12.24   & 56$\pm$16   & 6.05  & 3.3$\pm$0.9x10$^{-04}$ & - & - & - \\
	
8 & 03:38:41  & -35:31:35 & 10.75  & 1000$\pm$35 & 60.63 & 5.9$\pm$0.2x10$^{-03}$  & 99\% & y & - \\
	
9 & 03:38:50  & -35:24:57 & 12.24   & 47$\pm$14   & 5.02  & 2.8$\pm$0.9x10$^{-04}$ & - & - & - \\
	
10 & 03:39:08 & -35:31:22 & 12.24   & 56$\pm$16   & 6.19  & 3.3$\pm$0.9x10$^{-04}$ & - & - & - \\
	
11 & 03:38:32 & -35:27:05 & 13.70  & 171$\pm$30  & 11.53 & 1.0$\pm$0.2x10$^{-03}$  & - & - & - \\
	
12 & 03:38:52 & -35:26:42 & 11.62  & 295$\pm$22  & 24.66 & 1.7$\pm$0.1x10$^{-03}$  & 95\% & y & CGF 0102\\
	
13 & 03:38:49 & -35:28:35 & 12.24   & 77$\pm$20   & 7.68  & 4.5$\pm$1.1x10$^{-04}$ & - & y & - \\
	
14 & 03:39:10 & -35:27:07 & 12.24   & 46$\pm$14   & 5.06  & 2.8$\pm$0.8x10$^{-04}$ & - & y & - \\
	
15 & 03:38:02 & -35:26:27 & 17.32 & 62$\pm$19   & 4.70  & 3.6$\pm$1.1x10$^{-04}$   & - & - & - \\
	
16 & 03:38:29 & -35:25:05 & 17.32 & 75$\pm$23   & 4.82  & 4.4$\pm$1.3x10$^{-04}$   & - & - & - \\
	
17 & 03:38:32 & -35:30:57 & 12.24  & 70$\pm$18   & 6.81  & 4.1$\pm$1.1x10$^{-04}$  & - & - & - \\
	
18 & 03:38:04 & -35:33:27 & 17.32 & 65$\pm$19   & 5.38  & 4.1$\pm$1.2x10$^{-04}$   & - & y & CGF 0354\\
	
19 & 03:38:08 & -35:34:20 & 14.98 & 908$\pm$34  & 52.35 & 5.7$\pm$0.2x10$^{-03}$   & 99\% & y & - \\
	
20 & 03:38:43 & -35:33:50 & 15.19 & 105$\pm$18  & 8.67  & 6.3$\pm$1.1x10$^{-04}$   & - & y & - \\
\\
\hline

\end{supertabular}
\label{sources_tab}
\caption{List of the sources detected by the wavelets algorithm of Damiani et al.(\cite{Dam97a}, \cite{Dam97b}). The meaning of each column is explained in sec.\ref{sources} CGF=Hilker et al.(\cite{Hilk99}); FCCB=Smith et al.(\cite{Smith96}).}
\end{table}

\begin{table}[p]
\begin{tabular}{cccrrrcccc}
\hline
N. & R.A.         & Dec.         & 3$\sigma$ Radius & Counts & Max S/N & Count Rate & Time Var.& Opt.id. & NED obj.\\
& & & (arcsec) & & & (Counts/sec) & (KS test) & & \\
\hline
\\						   					
21 & 03:39:20 & -35:29:22 & 16.53 & 527$\pm$30   & 33.02  & 3.2$\pm$0.2x10$^{-03}$ & 99\% & y & - \\
 
22 & 03:38:24 & -35:14:52 & 17.32 & 60$\pm$18    & 4.94	& 3.8$\pm$1.2x10$^{-04}$   & - & y & - \\

23 & 03:38:12 & -35:16:17 & 17.32 & 59$\pm$18    & 4.85	& 3.7$\pm$1.1x10$^{-04}$   & - & y & - \\
	
24 & 03:37:59 & -35:19:42 & 16.72 & 379$\pm$27   & 24.93  & 2.3$\pm$0.2x10$^{-03}$ & - & y & - \\
 
25 & 03:37:35 & -35:24:04 & 17.32 & 64$\pm$19    & 5.32	& 4.0$\pm$1.2x10$^{-04}$   & - & - & - \\
	
26 & 03:38:43 & -35:33:10 & 17.32 & 73$\pm$21    & 5.62	& 4.3$\pm$1.2x10$^{-04}$   & - & y & - \\
	
29 & 03:39:43 & -35:24:11 & 24.27 & 143$\pm$27   & 8.57	& 1.0$\pm$0.2x10$^{-03}$   & - & y & - \\
	
30 & 03:38:31 & -35:14:17 & 26.37 & 586$\pm$38   & 26.55  & 3.8$\pm$0.2x10$^{-03}$ & 99\% & y & FCCB 1263\\
	
31 & 03:38:17 & -35:15:57 & 24.49 & 124$\pm$31   & 7.18	& 7.9$\pm$2.0x10$^{-04}$   & - & y & - \\
	
32 & 03:38:26 & -35:25:27 & 24.49 & 145$\pm$37   & 6.25	& 8.5$\pm$2.2x10$^{-04}$   & - & y & - \\
	
33 & 03:37:36 & -35:33:37 & 24.24 & 921$\pm$40   & 41.43  & 6.0$\pm$0.3x10$^{-03}$ & - & y & - \\
	
34 & 03:37:39 & -35:34:17 & 21.90 & 426$\pm$31   & 24.24  & 2.8$\pm$0.2x10$^{-03}$ & - & y & - \\
	
35 & 03:37:50 & -35:21:27 & 34.63 & 139$\pm$37   & 5.79	& 8.4$\pm$2.3x10$^{-04}$   & - & y & - \\
	
36 & 03:39:28 & -35:18:52 & 34.63 & 129$\pm$35   & 5.58	& 8.6$\pm$2.3x10$^{-04}$   & - & y & - \\
	
37 & 03:37:36 & -35:14:47 & 48.99 & 153$\pm$44   & 5.07	& 1.1$\pm$0.3x10$^{-03}$   & - & - & - \\
	
38 & 03:37:33 & -35:17:17 & 48.99 & 190$\pm$49   & 5.97	& 1.3$\pm$0.3x10$^{-03}$   & - & - & - \\
	
39 & 03:37:29 & -35:41:11 & 69.28 & 580$\pm$119  & 14.13  & 6.7$\pm$1.4x10$^{-03}$ & - & y & - \\
	
40 & 03:38:15 & -35:17:57 & 69.28 & 244$\pm$67   & 5.18	& 1.5$\pm$0.4x10$^{-03}$   & - & - & - \\
	
41 & 03:38:37 & -35:41:32 & 62.94 & 366$\pm$61   & 8.88	& 2.4$\pm$0.4x10$^{-03}$   & - & - & - \\
	
42 & 03:39:12 & -35:26:12 & 97.98 & 312$\pm$91   & 4.67	& 1.9$\pm$0.5x10$^{-03}$   & - & - & - \\
	
43 & 03:38:12 & -35:30:02 & 97.98 & 365$\pm$101  & 4.98	& 2.2$\pm$0.6x10$^{-03}$   & - & - & - \\
\\
\hline
\end{tabular}
\addtocounter{table}{-1}
\caption{- Continued.}
\end{table}

\end{landscape}

\clearpage

Even though the mean source counts across the field are compatible with the expected background counts, their spatial distribution is not homogeneous. In Figure \ref{src_excess} we measured the sources excess in circular annuli centered on NGC 1399. We find that the excess is peaked on the dominant galaxy suggesting that the inner sources are associated with NGC 1399, rather than being background objects. In fact, excluding the central 300'', the measured and expected counts turn out to be in agreement within less than $1\sigma$ ($204\pm32$ vs $230\pm15$) 

\section{Conclusions}
We presented some preliminary results obtained from the analysis of a deep observation of the cD galaxy NGC 1399. We found that:
\begin{itemize}
\item NGC 1399 posseses an extended and asymmetric X-ray halo with a  multi-component structure;
\item the ``inner'' component is centered on the optical galaxy and follows a Beta profile, very similar to the one seen in non-dominant galaxies (e.g. NGC 1404);
\item the ``outer'' component displays a more complex morphology, probably affected by the cluster potential and by the interaction with the intracluster medium;
\item the individual sources found in the HRI FOV are slightly in excess of the number expected from deep surveys. The excess is strongly peaked on NGC 1399 suggesting that it is due to sources associated with the central galaxy.
\end{itemize}
~\newline
We aknowledge partial support from the CXC contract NAS8-39073 and NASA grant NAG5-3584 (ADP),
and from the MURST PRIN-COFIN 1998-1999 (resp. G. Peres).\\
This research has made use of the NASA/IPAC Extragalactic Database (NED) which is operated by the Jet Propulsion Laboratory, California Institute of Technology, under contract with the National Aeronautics and Space Administration.

\end{document}